# ESL-PSC Toolkit: a graphical software environment for linking shared genetic changes to convergent phenotypes

John B. Allard[1,2*] and Sudhir Kumar[1,2]

*Corresponding author:
john.allard@temple.edu

Affiliations
1 Institute for Genomics and Evolutionary Medicine, Temple University, Philadelphia, PA 19122, USA
2 Department of Biology, Temple University, Philadelphia, PA 19122, USA

## Abstract

Convergent evolution provides a useful framework for testing whether independent origins of similar traits share common genetic mechanisms. Evolutionary Sparse Learning with Paired Species Contrast (ESL-PSC) is an approach to identify genes and sites associated with convergent traits from aligned sequences by fitting sparse predictive models to phylogenetically informed species contrasts. However, practical use of ESL-PSC currently requires substantial command-line fluency for data assembly, species-pair design, execution, and output interpretation. Here we present an integrated ESL-PSC analysis environment (ESL-PSC Toolkit) centered on a graphical user interface (GUI). ESL-PSC Toolkit is designed to assist users from experimental design through data interpretation without requiring extensive technical expertise. It supports guided input validation, interactive tree-based pair selection, command preview, live execution, post-run exploration of ranked genes and aligned sites, a complementary substitution-counting method, and analysis of continuous quantitative convergent traits. The computational backend has been reimplemented in Rust with many performance optimizations and parallelism, greatly reducing runtime for most analyses and enabling cross-platform packaged distributions. Downloadable GUI and CLI toolkit software packages for Mac, Windows, and Linux are available at https://github.com/John-Allard/ESL-PSC/releases/latest

## Introduction

Repeated evolution of similar traits in independent lineages offers a powerful opportunity to investigate whether similar genomic changes contributed to the origins of those traits (Stern 2013; Allard and Kumar 2026). This logic has motivated extensive searches for molecular convergence across diverse systems, but proteome-scale analyses remain difficult because candidate signals are sparse, background convergence can be substantial, and trait-relevant changes may be distributed across many genes rather than concentrated in a few obvious loci (Allard et al. 2025). These properties favor approaches that can detect weak but coordinated signal across many sites while suppressing large numbers of irrelevant features.

The Evolutionary Sparse Learning with Paired Species Contrast (ESL-PSC) approach can address some of these challenges (Allard et al. 2025). In ESL-PSC, trait-bearing and trait-negative species are organized into phylogenetically informed contrast pairs, and sparse-learning models are trained on aligned sequences to identify genes and sites associated with the trait. Sparse learning is useful in this setting because only a minority of genes and sites are expected to contribute meaningfully to a convergent phenotype, whereas most aligned positions represent background variation. Sparse-group regularization, therefore, allows the method to favor a small subset of informative genes and sites while still using signal distributed across many loci. The paired-species-contrast strategy is equally important. By restricting the analysis to trait-positive and trait-negative species linked by defensible phylogenetic contrasts, ESL-PSC focuses on repeated trait transitions, masks substitutions on shared ancestral branches, and reduces the impact of broad background divergence. Choosing those species pairs, however, is non-trivial. It requires biological and phylogenetic judgment about which taxa should be contrasted, which alternative pairings are defensible, and whether the resulting design is appropriate for the question being asked.

The broader ESL framework applies sparse machine learning directly to molecular phylogenomic data, and its implementation relies on sparse-group regularization that favors a small subset of informative genes and sites (Simon et al. 2013; Kumar and Sharma 2021). The ESL-PSC approach could recover biologically relevant signals in empirical analyses of C4 photosynthesis and echolocation and performed well in simulations (Allard et al. 2025). Those results established the analytical value of the method, but practical use still required substantial command-line fluency, careful manual construction of species-group files, and separate downstream interpretation steps. Here, we present a user-friendly software environment in which the graphical user interface (GUI) serves as a convenient entry point for most users, while an improved command-line toolkit is made available for large-scale automation and programmatic workflows.

## Results

### Interface Components

The ESL-PSC Toolkit GUI offers a guided workflow that covers input validation, parameter selection, execution, and post-run interpretation (Fig. 1A). Inputs that previously had to be assembled manually as command-line arguments are instead selected through dedicated interface controls, and tooltips explain the meaning of the major options and outputs while sensible defaults are already populated for common use cases. This reduces the initial barrier for new users because the available analysis settings can be inspected directly in the interface rather than inferred from command syntax alone.

*Tree Viewer.* The tree viewer is a central component because it moves experimental design onto the phylogeny itself (Fig. 1B). By loading a Newick tree, users can create a species-group design interactively rather than writing a text file of species identifiers by hand. Pairs can be assigned by clicking directly on taxa, phenotype annotations can be loaded and displayed by color for binary or continuous traits, and a paintbrush-style editing tool can be used to create or revise phenotype annotations within the same view. When phenotype information is available, the built-in pair-selection algorithms can also automatically generate contrast sets according to several user priorities, including deterministic, shortest-distance, longest-sequence, maximum-contrast, random, and composite strategies (Fig. 1C). For continuous phenotypes, users can define upper and lower thresholds for high- and low-value candidate species, or require a minimum local percentage shift from low to high when selecting contrast pairs (Fig. 1D,E). The viewer can also export an annotated SVG showing phenotypes and selected pairs, providing a convenient record of the design and a ready-made figure element for presentations or manuscripts.

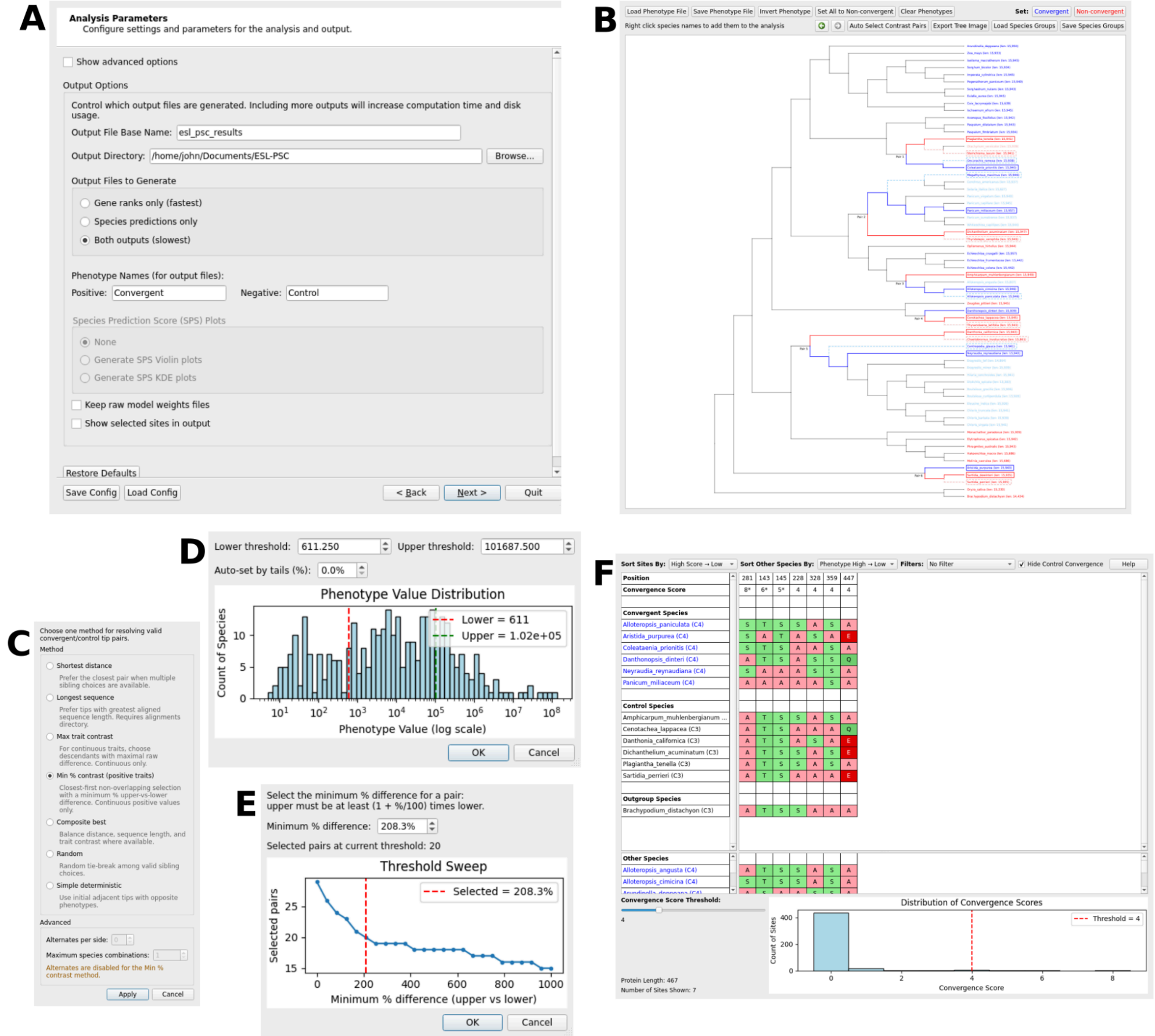

**Figure 1.** Major components of the ESL-PSC graphical interface. **A.** Analysis parameters are configured through a form-based interface with defaults and tooltips, allowing output settings and major run options to be specified without manual command assembly. **B.** The Tree Viewer displays the input phylogeny with phenotype annotations and selected paired-species contrasts, allowing users to construct or inspect an experimental design directly on the tree. **C.** Automatic pair-selection methods can be chosen from a dialog that presents multiple design strategies. **D** Continuous-phenotype thresholds can be defined interactively for pair selection. **E.** A threshold-sweep plot helps users choose a minimum percent-difference cutoff for local percent-contrast pairing. **F.** The Site Viewer provides residue-level interpretation by displaying aligned sequences with training species grouped by phenotype class, additional species shown separately below, and filters for convergence score and CCS-style site patterns.

## *Output displays*.

To demonstrate the updated software environment, we highlight representative outputs from both binary and continuous analyses together with a runtime benchmark. The new species-level Species Prediction Score (SPS) (Kumar and Sharma 2021) plot can be generated automatically from the photosynthesis example when phenotype annotations are available for the test species (Allard et al. 2025). Mean SPS values are plotted for the true C3 and C4 groups, providing a direct species-level summary of the prediction output without requiring the user to aggregate raw prediction rows separately. This output is closely aligned with the type of species-level interpretation used in the earlier ESL-PSC study, but it is now produced directly by the software as an optional output (Fig. 2A).

*Site Viewer.*

The Site Viewer provides the complementary residue-level interpretation layer (Fig. 1F). It functions as an alignment viewer specialized for convergence analyses and can also be opened independently for manual inspection of any alignment. Species used in the training pairs are grouped by phenotype class in the upper pane of the display so that residues distinguishing the modeled groups can be assessed immediately, whereas all remaining species are shown in a separate lower pane. When phenotype annotations are available, their labels are highlighted accordingly, allowing the broader distribution of residues to be scanned at a glance. Columns can be filtered either by a convergence-score threshold or by a CCS-style (Xu et al. 2017) site criterion, allowing the user to move quickly from ranked genes to explicit site patterns in the underlying alignment. Because this view combines phenotype-aware grouping with direct residue inspection, it is useful both for interpreting ESL-PSC results and for stand-alone manual review of candidate convergent sites.

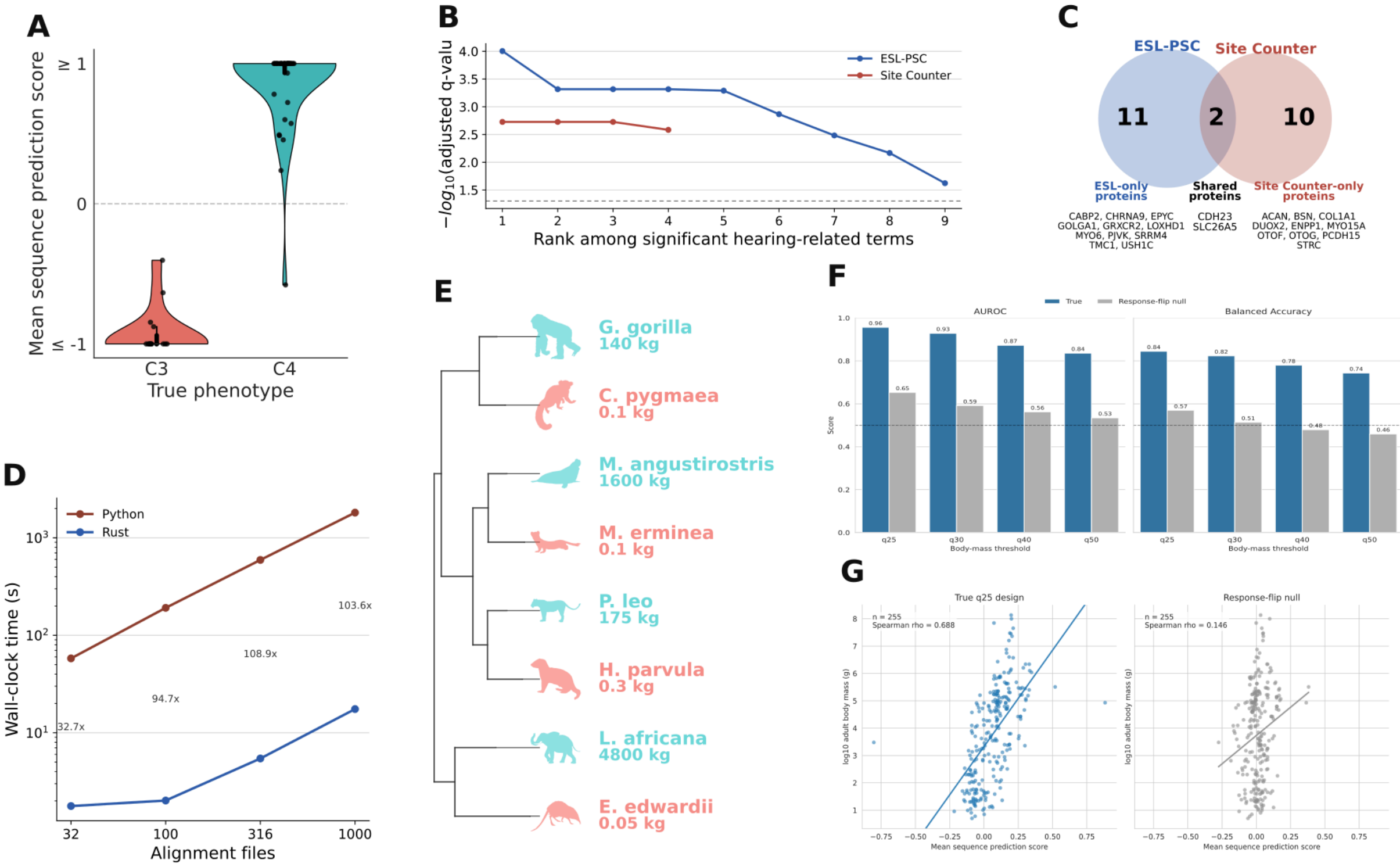

**Figure 2.** Representative ESL-PSC outputs, follow-up interpretation, continuous-trait analysis, and runtime performance. **A** Species-level prediction display from the photosynthesis demonstration. Violin distributions and overlaid species points show the resulting mean SPS values for C3 versus C4 taxa, with scores clipped at -1 and 1 for display. **B** Enrichment-significance profile for hearing-related terms from the echolocation example. For the echolocating-mammal analysis, the top-ranked ESL-PSC genes and the corresponding Site Counter output were each submitted to enrichment analysis, and terms related to hearing or auditory biology were extracted and ranked by adjusted q-value. The y-axis shows -log10(adjusted q-value), so larger values indicate stronger evidence of enrichment support. **C** Overlap of hearing-related proteins recovered by ESL-PSC and Site Counter in the echolocation example. The circles summarize the proteins contributing to the hearing-related enrichments in panel B, partitioned into proteins unique to the ESL-PSC ranking, shared between methods, and unique to Site Counter. **D** Runtime comparison between legacy Python-centered execution and the newer Rust-centered implementation. Wall-clock times are shown for benchmark runs using increasing numbers of alignment files as inputs. **E** Training design for the adult body-mass continuous-trait example. The eight training species used in the q25 body-mass analysis are arranged as four high-mass versus low-mass contrast pairs: *Gorilla gorilla* versus *Callithrix pygmaea*, *Mirounga angustirostris* versus *Mustela erminea*, *Panthera leo* versus *Helogale parvula*, and *Loxodonta africana* versus *Elephantulus edwardii*. Labels give abbreviated binomials and approximate adult body masses in kilograms. Silhouettes are from PhyloPic: *Gorilla gorilla gorilla*, T. Michael Keesey (after Colin M. L. Burnett), CC0 1.0; *Callithrix jacchus* proxy, Yan Wong from drawing by T. F. Zimmermann, CC0 1.0; *Mirounga leonina* proxy, Alexandre Vong, CC0 1.0; *Mustela erminea*, Margot Michaud, CC0 1.0; *Panthera leo*, Margot Michaud, CC0 1.0; *Helogale parvula*, Margot Michaud, CC0 1.0; *Loxodonta africana*, Guillaume Dera, CC0 1.0; *Rhynchocyon petersi* proxy, Kai Caspar, CC0 1.0. **F** Binary performance of the q25 adult body-mass model on independent held-out species. Evaluation was restricted to held-out species that were not descendants of the most recent common ancestor of any training pair. Performance is shown for the true design and for the built-in response-flip null model control. The left half reports AUROC, and the right half reports balanced accuracy at four phenotype binarization thresholds (q25, q30, q40, and q50), where species were labeled high or low body mass according to the corresponding global body-mass cutoffs, and middle species were not tested. **G** Continuous-trait prediction for the q25 adult body-mass example. The same species-averaged SPS values used in panel F are plotted against log10 adult body mass (grams) for all phylogenetically independent held-out species with body-mass data. The left subplot shows the true q25 design, and the right subplot shows the response-flip null. Spearman correlation is reported in each panel, and the fitted line is shown only to summarize the visual trend.

## Implementation and Analytical Capabilities

We describe the main implementation changes below and include representative analyses where they illustrate newly integrated toolkit capabilities.

### Software

ESL-PSC Toolkit is implemented as a cross-platform software package that includes a PySide6 desktop application, a terminal toolkit, and compiled analysis binaries. The toolkit is organized around a unified esl-psc command with subcommands for the main analysis, automatic pair selection, Site Counter, and plotting, so the same core operations are available from the GUI and the terminal. The GUI uses this shared command surface explicitly by generating the exact analysis command for the current configuration, which can be copied directly for scripted or batch execution. Prebuilt standalone GUI packages are distributed for Windows, macOS, and Linux, and the toolkit is packaged separately for terminal deployment, making the same analysis system available as both a desktop application and an automation-ready command-line environment.

### Analysis engine

The original ESL-PSC software (Allard et al. 2025) made use of the Sparse Group LASSO compute core from the MyESL package (Sanderford et al. 2025), which was based on the SLEP package (Liu et al. 2009). The principal analysis engine for ESL-PSC Toolkit has now been fully reimplemented in Rust and combines gap cancellation, preprocessing, sparse-group-lasso fitting over sparsity parameter combinations (lambda values), multi-species combination orchestration, and generation of ESL-style outputs within one compiled runtime (see Allard et al. (2025) for ESL-PSC pipeline details). This removes the older loop of repeated Python coordination, intermediate path files, preprocess-directory rewrites, and repeated solver input-output parsing. In the current implementation, each species combination can remain in memory while lambda values and group-penalty settings are swept, which reduces repeated disk traffic and data translation overhead. Model fitting is parallelized across lambda value combinations, which is a major source of the current runtime improvement on large alignment collections and dense grids of lambda value combinations. The execution framework also supports checkpointed multi-combination runs, allowing interrupted analyses to resume from completed species combinations instead of repeating the full analysis.

Finally, the runtime benchmark summarizes the effect of the new backend. Using the echolocation design with a 20 x 20 lambda grid and increasing alignment subsets, the parallel Rust implementation was faster than the legacy Python-centered path at every tested scale, with speedups of approximately 33-fold at 32 alignments, 95-fold at 100 alignments, 109-fold at 316 alignments, and 104-fold at 1000 alignments (Fig. 2D). The software expansion therefore improves not only accessibility and interpretability, but also the practical throughput needed for larger comparative analyses (Fig. 2D).

**Continuous-trait and pair-selection capabilities**

The current package also implements continuous-phenotype and pair-selection capabilities that were not previously unified in one environment. The Tree Viewer controls described above are backed by pair-selection workflows for both binary and quantitative phenotypes. For continuous phenotypes, users can define high- and low-value candidate species using upper and lower thresholds, or use a local percent-contrast selector for strictly positive traits. Tree-based pair selection supports deterministic, shortest-distance, longest-sequence, maximum-contrast, composite, random, and local percent-contrast strategies. The composite selector scores candidate pairings by standardized trait contrast, modulates that score by sequence support derived from the harmonic mean of paired sequence lengths, and resolves near-ties using phylogenetic structure. These capabilities allow quantitative phenotypes such as adult body mass, metabolic rate, age at maturity, or maximum longevity to be used in pair design and in phenotype-versus-SPS evaluation.

As a representative continuous-phenotype demonstration, we analyzed adult body mass in mammals (see Supplementary Methods). The training design was generated automatically by selecting candidate high- and low-mass species from the upper and lower quartiles of the phenotype distribution, then constructing four high-mass versus low-mass PSC-compliant contrast pairs with the pair-selection algorithm (Fig. 2E).

We then evaluated the resulting model only on species that were phylogenetically independent of the training pairs by excluding all descendants of the MRCA of each training pair from the test set. Under this criterion, the true model outperformed the response-flip null across AUROC and balanced-accuracy summaries at several threshold definitions used to binarize the evaluation species (Fig. 2F). The response-flip null did not collapse to AUROC = 0.5 because, with four training pairs, the balanced response-flip construction necessarily preserves two correctly oriented pairs in each null realization and therefore retains some convergent signal (Fig. 2F). The same body-mass example also showed a strong relationship between mean SPS and log10 adult body mass across held-out species, whereas the response-flip null retained only a weaker residual correlation (Fig. 2G).

**Site counter system.**

The package also includes Site Counter, a complementary substitution-counting workflow introduced to retain the direct interpretability of explicit convergent site patterns while improving on the limitations of a strict CCS-style screen (Xu et al. 2017). In the previous echolocation comparison, CCS was the most comparable counting-based approach, but its signal was limited and it did not benefit from the multiple valid species combinations available for the analysis. Site Counter addresses this by using the same multi-combination species design used by ESL-PSC and then aggregating gene-level evidence across combinations rather than relying on a single fixed pairing scheme. It also supports both fixed-outgroup and tree-based ancestral-state modes. In ancestral mode, MRCA states are reconstructed by parsimony, ambiguous ancestral states can be excluded, and the requirement that all control species exactly match the outgroup or ancestral residue can be relaxed through a configurable minimum control-agreement threshold. This produces a counting-based companion method that is more flexible than CCS while remaining closely tied to explicit site patterns in the alignment.

We compared the main ESL-PSC ranking with Site Counter, the complementary multi-combination substitution-counting workflow, on the echolocation example using the same 16 species combinations and the top 100 ranked proteins from each method. Restricting the display to significant hearing-related enrichments, ESL-PSC recovered nine significant terms, whereas Site Counter recovered four, and the strongest ESL-PSC enrichments also had better adjusted q-values overall (Fig. 2B). Site Counter nevertheless recovered significant hearing-related terms, indicating that the complementary counting-based method retains biologically relevant signal even though it is less information-rich than the full ESL-PSC modeling output (Fig. 2B). This is noteworthy in light of the earlier ESL-PSC comparison with CCS, in which the counting-based approach recovered only limited echolocation signal and overlapped minimally with the ESL-PSC top-ranking proteins. Site Counter improves on that earlier counting-style strategy by aggregating evidence across multiple species combinations and by relaxing the control-versus-outgroup matching rule when appropriate. Site Counter also produced additional significant non-hearing enrichments that are not shown in the figure, consistent with the broader behavior of CCS-like screens in which relevant and non-target signals can both emerge from the same ranked output (Fig. 2B).

The overlap analysis shows that the two methods are complementary rather than redundant. The proteins contributing to the significant hearing-related enrichments were mostly not shared between methods. In the current example, ESL-PSC contributed 13 hearing-related proteins, Site Counter contributed 12, and only two proteins were shared. Thus, the full ESL-PSC model and the complementary substitution-counting method do not simply return the same candidates at different speeds; rather, they identify partly distinct protein sets that converge on related biological interpretation (Fig. 2C).

## Discussion

The main contribution of the present software is to make ESL-PSC more usable without reducing its analytical depth. Some previous analyses established that supervised sparse learning with paired species contrasts can recover biologically meaningful signals of convergent sequence evolution (Allard et al. 2025). The current software environment makes that framework easier to apply by integrating experiment setup, phylogenetic design, execution, complementary screening, and output interpretation in one coherent workflow. Such integration and user-friendliness matter especially for convergence research because many of the most interesting systems are studied by researchers whose expertise lies in organismal biology, trait evolution, physiology, or natural history rather than in computational pipeline construction and validation. Those researchers are often best positioned to define meaningful comparative questions, but historically they have faced a substantial technical barrier in applying command-line implementations of sophisticated methods (Kumar and Dudley 2007). By lowering that barrier while preserving methodological transparency, the present software should make it easier for a broader community of domain experts to test convergent hypotheses directly in the biological systems they know best.

At the same time, the command-line toolkit remains important. It is well-suited to large-scale automation, repeated runs across many designs, and AI-assisted workflows in which analyses are generated, launched, and summarized programmatically. The GUI and CLI should therefore be viewed as complementary interfaces to the same platform rather than as competing alternatives. More broadly, the software now combines tree-based design, continuous trait support, complementary substitution counting, residue-level interpretation, cross-platform packaging, and a parallel high-performance backend. ESL-PSC has thus evolved from a method implementation into a mature software environment for convergent sequence analysis.

**Data Availability**

Source code is available at https://github.com/John-Allard/ESL-PSC. Downloadable GUI and CLI toolkit software packages for Mac, Windows, and Linux are available at https://github.com/John-Allard/ESL-PSC/releases/latest.